\documentstyle[12pt,twoside,fleqn,espcrc1,epsf]{article}

\def\Journal#1#2#3#4{{#1} {\bf #2}, #3 (#4)}

\def\NPA{{\em Nucl. Phys.} A}

\def\PRL{\em Phys. Rev. Lett.}
\def\PRD{{\em Phys. Rev.} D}
\def\PRC{{\em Phys. Rev.} C}
\def\ZPC{{\em Z. Phys.} C}
\def\PR{{\em Phys. Rep.}}
\def\CPC{{\em Comp. Phys. Comm.}}
\def\MPLA{{\em Mod. Phys. Lett.} A}

\newcommand{\AmS}{{\protect\the\textfont2
  A\kern-.1667em\lower.5ex\hbox{M}\kern-.125emS}}

\title{A Parton-Hadron Cascade Approach in High-energy Nuclear Collisions}

\author{Yasushi Nara\address{Advanced Science Research Center, 
        Japan Atomic Energy Research Institute, \\
        Tokai, Naka, Ibaraki 319-11, Japan}}

\begin{document}
\maketitle

\begin{abstract}
 A parton-hadron cascade model which is the extension of hadronic
 cascade model incorporating hard partonic scattering
  based on HIJING is presented to describe the
space-time evolution of parton/hadron system produced by
ultra-relativistic nuclear collisions.
Hadron yield, baryon stopping and transverse momentum distribution
are calculated and compared with HIJING and VNI.
\end{abstract}

\section{INTRODUCTION}
Event generators based on perturbative QCD (pQCD) are proposed such as
HIJING (Heavy Ion Jet Interaction Generator)\cite{hijing},
VNI (Vincent Le Cucurullo Con Giginello)\cite{vni},
in order to describe ultra-relativistic heavy ion collisions,
  especially, at collider energies (RHIC/LHC).
Although both models reproduce many  $pp$ and ${\bar p}p$ data,
there are large discrepancies in AA collisions between these two models,
for example, absolute number of produced particles,
slope of transverse momentum.
The main differences of the modeling between these two models are following:
VNI is a Monte Carlo implementation of parton cascade model (PCM)
in which the time evolution of heavy ion collision is simulated
by the parton cascading.
While HIJING assumes the Glauber theory in the description of
AA collisions and  handles the soft process based on the string model.
Namely, the treatment of the multiple parton and hadron interactions
is different.
Second,
the cut-off $p_0$ has to be introduced to avoid divergent QCD cross section
for $p_{\bot} \to 0$ and this value is model dependent.
In order to understand particle production mechanism in nuclear collisions,
the role of multiple interactions as well as the sensitivity of model parameters
should be carefully investigated.
Hadronic microscopic transport models such as
RQMD~\cite{sorge1} ARC~\cite{ARC}, QGSM~\cite{QGSM} and UrQMD~\cite{UrQMD}
 have been successfully applied to nuclear reactions
at AGS and SPS energies.
Thus, the purpose of this work is to
 include  hard processes into hadronic transport model
 by using HIJING formalism
 and study the effect of multi-step interaction
 on particle spectra at collider energies.

\section{PARTON-HADRON CASCADE MODEL}
In parton-hadron cascade model (PHC),
elementary processes are taken from HIJING;
The hard and soft processes are determined
 by HIJING formalism\cite{xnwang1} and PTHIA5.7\cite{pythia57} is used to
generate hard scattering as well as initial and final state
radiations.
Only on-shell partons produced from hard scattering are propagated.
For soft interaction of hadrons,
the string excitation is assumed
  with the same probability for light-cone momentum exchange
  as DPM type functions at the c.m. energy above 5GeV.
 At low energy ($\sqrt{s}<5$GeV), $1/x$ distribution
  which is the same as FRITIOF~\cite{fritiof} model is used.
If excited mass is larger than 2GeV in baryon case,
 it is assumed to be excited string like.
The strings are assumed to hadronize via quark-antiquark creation
using Lund fragmentation subroutine PYSTRF of PYTHIA6.1\cite{pythia57}.
Therefore, at hh level PHC is essentially the same as HIJING.

For the description of AA collisions,
 the trajectories of all hadrons as well as partons,
 including produced particles, are followed explicitly
   as a function of space and time.
Space-time point of produced partons can be simply determined by
 the uncertainty principle\cite{eskola1}.
Formation points of hadrons from the fragmentation of string
are assumed to be fixed by
  the average of two constituents formation points\cite{sorge1}.
 Low energy baryon-baryon, baryon-meson and  meson-meson rescattering
 are also included assuming resonance excitation picture
 in order to treat final state interaction of hadronic gas.
Extending the particle table of PYTHIA, baryon and meson resonances
are explicitly propagated and they can rescatter.
The rescattering among produced partons are not implemented now,
however, constituent quarks can scatter with hadrons
  assuming the additive quark cross section within a formation time.
The importance of this quark(diquark)-hadron interaction
 for the description of baryon stopping
at CERN/SPS energies was reported by Frankfurt group~\cite{sorge2,UrQMD}.

\begin{figure}[t]
  \centerline{\epsfxsize=16cm \epsfbox{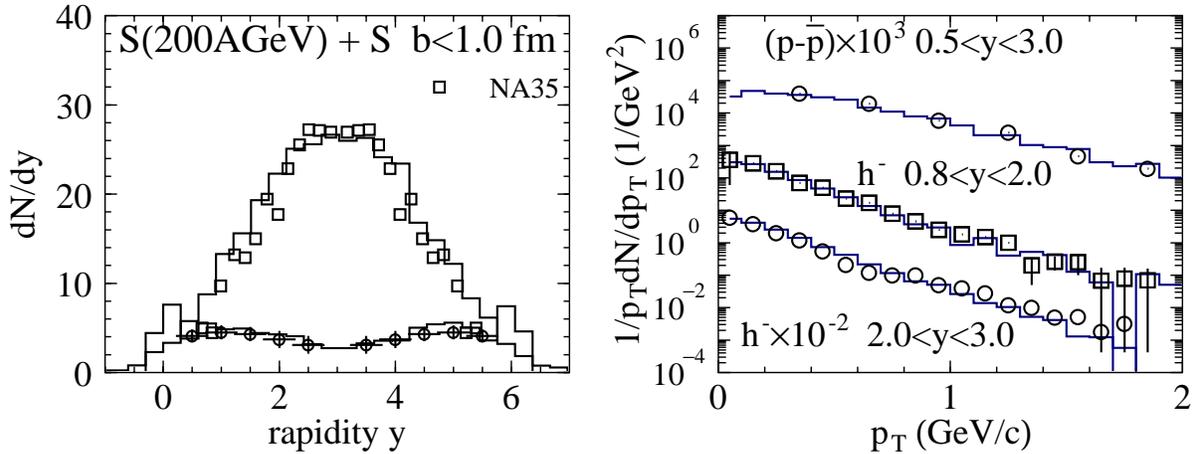}}
  \caption{The rapidity and transverse momentum distributions of net protons
        and negative charged particles ($\pi^-,K^-,{\bar p}$)
        for S + S collision at 200GeV/c with centrality 20\%.
       Experimental data are taken from NA35. 
          }\label{fig:ss}
\end{figure}

\section{RESULTS}

In Fig~\ref{fig:ss},
I compare the data~\cite{na35} on net proton and negative charged particle
 rapidity and transverse distributions
 for S+S collision at 200AGeV/c.
PHC improves the HIJING results~\cite{gyulassy} and
the agreement is good for both net protons and negative charged particles.
At this energy, this model is reduced to the hadronic cascade model
 if we use the cut off parameter $p_{\bot}=2$GeV,
 because the probability for hard scattering is very small
  at this incident energy
 ( average number of hard scattering $\sim$ 0.1/event in S + S collision ).
This model can work also well for the lower energies.

\begin{figure}[tb]
   \centerline{\epsfxsize=16cm \epsfbox{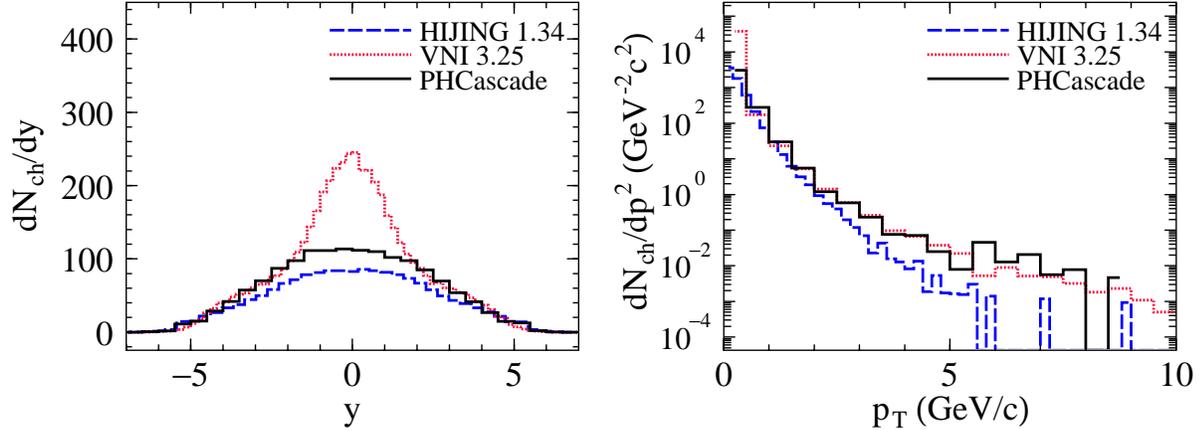}}
    \caption{The rapidity and transverse momentum distributions of
                charged particles for S + S collision
               at RHIC energy ($b=0.0$fm).
             Solid, long dashed and dots histogram represent
              PHC, HIJING and VNI results respectively.
             } \label{fig:rap}
\end{figure}

Figure~\ref{fig:rap} shows
   the charged particle rapidity and transverse momentum distributions from
   three different models, VNI, HIJING and present model
  in S+S collision at RHIC energy.
It is seen that 
 absolute particle yield in PHC calculation is not so different from HIJING,
 however, PHC result for the transverse momentum distribution becomes close to 
  the parton cascade model prediction. 
The difference between HIJING and PHC results comes mainly from
 the treatment of multiple collisions of hadrons and quarks.

\begin{figure}
   \centerline{\epsfxsize=9cm  \epsfbox{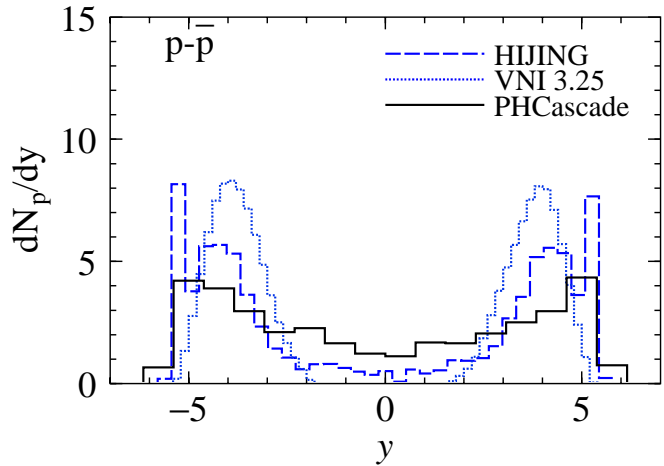}}
    \caption{The rapidity distributions of net proton for S +S collision
              at RHIC energy.
             Solid, long dashed and dots histogram represent
              PHC, HIJING and VNI results respectively.
             } \label{fig:rapp}
\end{figure}

The baryon stopping problem is one of the important element
in nucleus-nucleus collisions.
Net proton distributions calculated by various models
are compared in Fig.~\ref{fig:rapp}.
PHC model predicts no baryon free region at midrapidity,
 while parton cascade model predicts transparency.
PHC result is similiar to RQMD and UrQMD calculations~\cite{rqmd2,urqmd2}.
It is important to mention that
  if quark-hadron interaction is switched off,
PHC gives the same results as HIJING.
Interaction of quarks with hadrons modifies the stopping power as well as
transverse momentum shape.
What is the important difference between VNI and PHC is that
 first NN collisions are the scattering of the coherent objects (hadron)
 in PHC and after the hard scattering, partons are treated as classical
 particle like VNI,
 while in VNI,
    from the beginning,
    partons are sampled and
  on-shell as well as virtual partons are evolved in time.

In summary, a microscopic transport model which is the extension of
hadronic cascade model based on HIJING is presented.
The multi-step interaction in AA collisions changes
   the prediction for the final hadron distribution.
It is interesting to include rescattering between produced partons
to investigate thermalization and equilibration of parton system.
This may be also useful for the understanding of the
discrepancy between models.

\section*{ACKNOWLEDGMENTS}
I would like to thank Dr. A. Ohnishi
  for his encouragements and useful comments.
I acknowledge useful comments by Prof. M. Gyulassy.


\end{document}